\documentclass[aps,prd,11pt,preprint,nofootinbib,showkeys,longbibliograohy]{revtex4-2}
\usepackage{amsmath,amssymb,amsthm,mathrsfs,bbm}
\usepackage{latexsym,amscd,amsbsy,amsfonts,dsfont}
\usepackage{graphicx}
\usepackage[utf8]{inputenc}
\usepackage{subfigure}  
\usepackage{float}
\usepackage{slashed}
\usepackage{cancel}
\usepackage{multirow}
\usepackage{soul}
\usepackage{physics}
\usepackage{comment}
\usepackage{enumitem}
\usepackage[usenames,dvipsnames]{xcolor}
\usepackage[
      colorlinks=true,
      linktocpage=true,
      breaklinks=true,
      linkcolor=Aquamarine,
      urlcolor=Purple,
      filecolor=black,
      citecolor=Purple,
      menucolor=black,
      pdfstartview=FitV,
      bookmarksopen=true
      ]{hyperref}

\makeindex

\begin{document}

\title{The harmonic background paradigm, or why gravity is attractive}

\author{Carlos Barcel\'o}
\email{carlos@iaa.es}
\affiliation{Instituto de Astrof\'{\i}sica de Andaluc\'{\i}a (IAA-CSIC), Glorieta de la Astronom\'{\i}a, 18008 Granada, Spain}
\author{Gil Jannes}
\email{gjannes@ucm.es}
\affiliation{Department of Financial and Actuarial Economics \& Statistics, Universidad Complutense de Madrid, Campus Somosaguas s/n, 28223 Pozuelo de Alarc\'on (Madrid), Spain}

\begin{abstract}
In a work by Visser, Bassett and Liberati (VBL) [Nucl. Phys. B (Proc. Suppl.) 88, 267 (2000)] a relation was suggested between a null energy condition and the censorship of superluminal behaviour. Their result was soon challenged by Gao and Wald [Class. Quantum Grav. 17, 4999, (2000)] who argued that this relation is gauge dependent and therefore lacks physical significance. In this paper, we clear up this controversy by showing that both papers are correct but need to be interpreted in distinct paradigms. 
In this context, we introduce a new paradigm to interpret gravitational phenomena, which we call the Harmonic Background Paradigm. This harmonic background paradigm starts from the idea that there exists a more fundamental background causality provided by a flat spacetime geometry. One of the consequences of this paradigm is that the VBL relation provides an explanation of why gravity is attractive in all standard weak-field situations.
\end{abstract}

\keywords{}

\maketitle
 
\tableofcontents

\section{Introduction}
\label{Sec:Introduction}

Since at least the works of Rosen~\cite{Rosen1940a,Rosen1940b}, it is known that General Relativity (GR) can be interpreted from the point of view of two distinct paradigms. One is the so-called {\em Geometric Paradigm} (GP), which is the most standard manner in which General Relativity is presented and understood. The other can be called {\em Field-Theoretic Paradigm} (FTP) and corresponds to understanding GR as a non-linear, relativistic gauge theory of a tensorial field over a flat background. The latter paradigm is adopted in the description of all the other fundamental interactions found in Nature~(for a description of these paradigms, see e.g.~\cite{Cao2019}). 
We call these two ``paradigms'', instead of just formulations, because we think they affect one's view on gravitational phenomena and how to transcend the GR regime. 
Nonetheless, the literature shows no clear consensus on whether these paradigms as applied to GR are equivalent or on the contrary crucially distinct (for an historical overview of different positions, see~\cite{PittsSchieve2001}). 
For instance, an interesting and sharp but not very well-known result about the non-equivalence of these two paradigms was presented by Penrose~\cite{Penrose1980}.
In this paper he showed that the conformal (or causal) asymptotic structures of Minkowski spacetime and the simplest physical geometries appearing in GR do not match together.

The first central theme of this work is to describe and discuss what we argue to be a third paradigm in which to frame the gravitational interaction. We shall call this third paradigm the \emph{Harmonic Background Paradigm} for reasons we will explain later. This harmonic background paradigm is closer to the field-theoretic paradigm, but significantly differs from it essentially in that one imposes a specific physical gauge (in our perspective this is however not at all a gauge fixing). Similar ideas have been contemplated from time to time, starting from Rosen himself~\cite{Rosen1940a}, over the {\em Relativistic Theory of Gravitation} by Logunov and collaborators~\cite{LogunovMestvirishvili1985} up to the {\em Special Relativistic Approach} by Pitts \& Schieve~\cite{PittsSchieve2001}.
What these approaches have in common is to ascribe a crucial role to a background geometric structure. However, as these names might evoke entire sets of results, many of them different from what we are concretely proposing here, we have opted to use a different name. 
For the purpose of this introduction, let it suffice to say that, for most gravitational phenomena, these paradigms are physically identical and can be considered as just different interpretations of a single theory: General Relativity. However, the relevance of having different interpretations or paradigms is that they suggest rather different extensions beyond the current standard theory.

The other central theme of this work is a subtle connection between superluminality and energy condition violations put forward by Visser-Bassett-Liberati (VBL) in~\cite{VisserBassettLiberati2000}. 
This theme might at first seem unrelated to the first central theme stated previously. But, as the reader will see in due time, they are on the contrary tightly related. Let us briefly describe this second theme.

In GR, by construction, there is nothing that can travel faster than the local speed of light. This is because GR precisely prescribes at each spacetime point a causal-cone structure; the metric is nothing more than this causal-cone or conformal structure multiplied by a conformal or volume factor. 
This observation does however not completely eliminate the possibility of considering faster-than-light configurations. Such behaviours typically involve some non-local definition of superluminal behaviour. For example, in cosmology we can have geometries with cosmological horizons, in which space is expanding so rapidly (exponentially) that light rays cannot connect points which are initially sufficiently far apart. Another paradigmatic example are warp drive configurations~\cite{Alcubierre1994}.

All examples we know of superluminal behaviour within standard GR are based on configurations in which the causal-cones stretch more open in some regions of spacetime with respect to some other referential causal-cones in other regions. In general, it turns out that generating such configurations requires introducing some exotic matter content: The matter stress-energy tensor has to violate some of the energy conditions (EC) of GR~\cite{HawkingEllis1973}.
Armed with this intuition, and starting with \cite{Olum1998}, several attempts have been made to prove general theorems which make an explicit connection between EC violations and superluminal behaviour.
Among these results there is the one due to VBL that links EC satisfaction with the impossibility of travelling faster than a referential speed of light~\cite{VisserBassettLiberati2000}. However, this result was soon challenged by Gao and Wald~\cite{GaoWald2000}, who strongly criticised the VBL approach. Gao and Wald agree that there should exist connections between superluminality and EC violations, but they argue that such connections should be built using a different strategy, and in particular an explicitly gauge-invariant strategy. For instance, they go over and construct two theorems establishing links between superluminality and EC violations.

Both the VBL paper and the Gao-Wald response have (separately) received a fair amount of attention. In the VBL case, these range from warp drives~\cite{Alcubierreetal2017} and wormholes~\cite{Simpson2019} over extensions of the energy conditions~\cite{Bousso2016} to more general considerations of null cones~\cite{PittsSchieve2004} and causality~\cite{hertzberg2017general}. The Gao-Wald is also cited in the context of causality \cite{Camanho2016causality}, its relation to energy conditions \cite{Hartmanetal2017averaged} and their impact on IR/UV physics \cite{Alberteetal2022reverse}, but has furthermore found its way to discussions on holography \cite{Bousso2002holographic} and the (A)dS-CFT correspondence \cite{Strominger2001ds,Hubeny2015ads}. Curiously, the actual controversy between the VBL and Gao-Wald arguments has, to our knowledge, never been analyzed in detail.

The second central point of the present paper is thus to clarify this controversy. As we will argue, both arguments are in fact correct but they require the assumption of different conceptual setups or paradigms. It is precisely in making the VBL approach rigorously correct that we will find one motivation for introducing the harmonic background paradigm. 
We will argue that this not only clears up the VBL/Gao-Wald debate, but leads to many other interesting consequences. For instance, in this paper we contend that ``inverting'' the VBL interpretation of their own result offers a potential explanation of why gravity is attractive, at least under gravitational-weak-field conditions.

The structure of the paper is the following. In sections \ref{Sec:VBL} and \ref{Sec:GW} we review the VBL result and the Gao-Wald criticism, respectively. Then, we present what we call the harmonic background paradigm as a setup for discussing gravitational phenomena, in Section~\ref{Sec:ThirdP}. The discussion in this section also serves, among other things, to shed light onto the previous controversy. After that, we turn in section \ref{Sec:Why-attractive} to expose one of the central results of this work: we explain why gravity is attractive in standard weak-field situations. Section~\ref{Sec:Penrose} is devoted to a discussion of a potential obstruction to our proposal identified by Penrose in~\cite{Penrose1980}. Finally, we summarize our proposal and make further remarks in Section~\ref{Sec:Summary}.

\section{The Visser-Bassett-Liberati argument}
\label{Sec:VBL}

Let us suppose at this stage that VBL assume a standard geometric paradigm in their discussion. We assume this because in their paper there is no explicit discussion of a need for any paradigm shift. As we will argue later, however, in our opinion, they are making a deep observation at a heuristic level which deserves further elaboration. 

Presented in steps, the VBL result is essentially the following. 

\begin{enumerate}
\item
Whenever we have a metric $g_{\mu\nu}$ which is close to being flat we can write $g_{\mu\nu} \simeq \eta_{\mu\nu} + h_{\mu\nu}$, where $\eta_{\mu\nu} $ is the Minkowski metric. 

\item
From the Fierz-Pauli Lagrangian for linearized gravity~\cite{FierzPauli} one obtains the following equation for $h_{\mu\nu}$:
\begin{eqnarray}
\square h_{\mu\nu} 
- 2\eta_{\beta (\mu} \nabla_{\nu)} \nabla_{\alpha} h^{\alpha\beta}
+\left( \nabla_{\mu} \nabla_{\nu} h + \eta_{\mu\nu}  \nabla_{\alpha} \nabla_{\beta} h^{\alpha\beta} \right). 
-\eta_{\mu\nu} \square h = -16\pi T_{ab}
\label{Eq:FP}
\end{eqnarray}
Now, in the de Donder or harmonic gauge~\cite{dedonder-gravifique}\footnote{None of De Donder's works on relativity have, as far as we are aware, ever been translated in English. For a rare work of De Donder in English, see his lecture notes \cite{dedonder-course}.} 
\begin{eqnarray}
\nabla_\mu \left(h^{\mu\nu}-\frac{1}{2}\eta^{\mu\nu}h \right)=0,
\label{Eq:DonderGauge}
\end{eqnarray}
the Einstein equations read
\begin{eqnarray}
\square h_{\mu\nu} = -16\pi \bar{T}_{\mu\nu};~~~~~\bar{T}_{\mu\nu} = T_{\mu\nu} -{1 \over 2}\eta_{\mu\nu} T.
\label{Eq:LEEs}
\end{eqnarray}
In all the previous expressions, $\nabla$ represents the covariant derivative with respect to the background metric $\eta$, and we have used $\nabla$ instead of $\partial$ (as in the VBL paper) to stress the invariance of the construction under changes of coordinates.

\item
In the VBL paper, the next condition imposed is that the system does not contain any incoming gravitational waves, or equivalently, that there are no free gravitational waves superposed to the inhomogenous solutions of the D'Alembertian equation for $h_{\mu\nu}$. Once this condition is imposed, they assert that one can write down the further evolution of the perturbation $h_{\mu\nu}$ as 
\begin{eqnarray}
h_{\mu\nu}(\mathbf{x},t)= 16\pi\int d^3 y \frac{\bar{T}_{\mu\nu}(\mathbf{y},t')}{\abs{\mathbf{x}-\mathbf{y}}}.
\label{VBL-int}
\end{eqnarray}
The $\bar{T}_{\mu\nu}$ in the above equation depends on $\mathbf{y}$ and the retarded time $t'=t-\abs{\mathbf{x}-\mathbf{y}}/c$.

\item Finally, they assume that matter satisfies a Background Null Energy Condition or BNEC, namely that $T_{\mu\nu} l^\mu l^\nu \geq 0$ for any null vector $l^\mu$ with respect to $\eta$, i.e.\ any $l^\mu$ such that $\eta_{\mu\nu}l^\mu l^\nu=0$. Then, under the previous conditions they show that the causal cones of the $g$ metric will always lie inside the cones of the $\eta$ metric. The argument is simple: contracting \eqref{VBL-int} with $l^\mu l^\nu$ one sees that
\begin{eqnarray}
h_{\mu\nu}l^\mu\l^\nu \geq 0
\label{VBL-null}
\end{eqnarray}
But if $l^\mu$ is null with respect to $\eta$, then \eqref{VBL-null} is also the norm of $l^\mu$ with respect to the full metric $g$. Hence, if $l^\mu$ was null with respect to $\eta$, it can never be timelike, and in fact, unless $T_{\mu\nu}\equiv 0$, it will necessarily be spacelike, with respect to $g$. In other words, the causal cones of the $g$ metric always lie inside the cones of the $\eta$ metric. In the following we will write symbolically $[g]<[\eta]$. Note that the BNEC is not the standard NEC condition, since the latter would involve null vectors with respect to $g$ rather than $\eta$. 
\end{enumerate}

Thus, the essence of the VBL argument is the observation that, if the BNEC is satisfied, the $g$-lightcones will necessarily become squeezed inside the $\eta$-lightcones, and can never lie outside them (the proof they provide is only at the linear regime). That is why they called their result a ``superluminal censorship''.

\subsection{Some further subtleties}

Before continuing with the Gao-Wald counterargument, let us point out some subtleties of the VBL argument that were not explicitly discussed in the VBL paper. Before discussing the linear gravitational case, it is interesting to recall a very standard way of calculating the electromagnetic field associated, for instance, with an arbitrarily moving point-like charge. In the standard treatment of electromagnetism as a gauge theory, we know that the physical electric and magnetic fields do not depend on the gauge used to calculate them in terms of four-potentials. Nevertheless, it is typically useful to adopt some particular gauge to perform the calculations. One standard path is to first assume the Lorenz gauge $\nabla_\mu A^\mu=0$~\footnote{The Lorenz gauge is often mistakenly attributed to Lorentz, see Jackson \& Okun \cite{jackson2001historical} for an extensive discussion. Jackson's own classic textbook \cite{Jackson} in fact commits exactly this confusion up to the second edition. The third edition corrects the mistake and contains a note of explanation, but ironically still commits the misspelling in a few places.}. The four-potential then has to satisfy the D'Alembertian equation $\square A_\mu=0$. The solution is easily found in any textbook, e.g.~\cite{Jackson}, namely     
\begin{eqnarray}
A_\mu = A^{\rm in}_\mu + \int dx' G_{{\rm R}\mu}^{~~\mu'}(x,x') j_{\mu'}(x')
\end{eqnarray}
where $G_{\rm R}(x,x')$ is the retarded Green function
\begin{eqnarray}
G_{{\rm R}\mu}^{~~\mu'}(x,x')= \delta_{\mu}^{\mu'} {1 \over |\mathbf{x}-\mathbf{x}'|}\delta(t'-t+|\mathbf{x}-\mathbf{x}'|/c).
\end{eqnarray}
This specific Green function is selected by imposing specific boundary conditions or a concrete integration contour if one is working in the Fourier space~\cite{Griffel2002}. Here, to give a completely explicit expression, we have used Cartesian coordinates in the Minkowski spacetime in which the theory is defined. 
Then, for the example of a single point-like charge $j^\mu=q \gamma^{-1} u^\mu \delta(x-x_{\rm T}(t))$, one obtains the Lienard-Wiechert retarded potentials
\begin{eqnarray}
A^{\mu}_{\rm LW}(t,\mathbf{x}) = {q \gamma^{-1} u^\mu \over \kappa R}\bigg|_{\rm Ret}. 
\end{eqnarray}
In this expression $x_{\rm T}(t)$ is the trajectory of the charge, $u^\mu=\gamma\{1,\boldsymbol{\beta} \}$ its normalized four velocity, $\gamma$ the Lorentz factor, and $\kappa=1-\widehat{\mathbf{R}} \cdot \boldsymbol{\beta}$ with $\widehat{\mathbf{R}}$ a unit vector with direction between the point $\mathbf{x}$ and the retarded position of the charge. Then, the total potential is
\begin{eqnarray}
A_\mu = A^{\rm in}_\mu + A^{\rm LW}_{\mu}.
\end{eqnarray}
Having selected a unique Green function, the non-uniqueness of the solution to the D'Alembertian equation $\square A_\mu=0$ is encoded in the presence of the term $A^{\rm in}_\mu$ which represents any free solution of the equation.

Now, to calculate the field produced just by the moving charge it is customary to take  $A^{\rm in}_\mu=0$. Here we want to emphasize that setting $A^{\rm in}_\mu=0$ selects a unique four-potential to represent the physical situation, that is, completely fixes the initial gauge freedom for such situations.

In a more detailed manner, we could have written $A^{\rm in}_\mu$ as consisting of two parts:     
\begin{eqnarray}
A^{\rm in}_\mu = A^{\rm in-phys}_\mu + A^{\rm in-gauge}_\mu.
\end{eqnarray}
The first part embodies the presence of physically real waves incoming from infinity (i.e.\ waves with non-trivial electric and magnetic fields). 
The second part, the ``residual gauge'', is characterized by changes of $\delta A_\mu= \partial_\mu \phi$ such that $\square \phi=0$, and hence represents pure gauge waves. Imposing boundary conditions so that both terms are zero not only eliminates the physical incoming waves, but also fixes completely the gauge used in the calculation, including the residual freedom left by the choice of the Lorenz gauge. This second issue is typically not made explicit, probably because it is not important for the calculation at hand. Indeed, any consistent choice of $A^{\rm in-gauge}_\mu$ result into a zero contribution to the relevant electric and magnetic fields anyway. Notice also that fixing completely the gauge can not be attained using only local conditions on the 4-potential. We have needed in addition to impose some asymptotic boundary conditions.

The situation in the linear gravitational case is parallel but a bit more subtle. 
If one sticks to the geometric paradigm, and thus adheres to the idea that there is just one real, physical metric tensor $g_{\mu\nu}$, then an immediate question that arises is how to make the separation $g_{\mu\nu}=\eta_{\mu\nu}+ h_{\mu\nu}$. This separation is not unique, in fact a priori there are infinitely many possible such separations.
A first condition in the VBL analysis that restricts this set of partitions consists in imposing the harmonic condition (\ref{Eq:DonderGauge}) on $h_{\mu\nu}$. This condition allows to write down the general solution for $h_{\mu\nu}$ as 
\begin{eqnarray}
h_{\mu\nu}= h^{\rm in}_{\mu\nu} + 16\pi\int d^3 y \frac{\bar{T}_{\mu\nu}}{\abs{\mathbf{x}-\mathbf{y}}}.
\label{VBL-int2}
\end{eqnarray}
where $h^{\rm in}_{\mu\nu}$ are free solutions of the d'Alembertian equation $\square h_{\mu\nu}=0$.

The first term on the right-hand side is 
\begin{eqnarray}
h^{\rm in}_{\mu\nu} = h^{\rm in-phys}_{\mu\nu} + h^{\rm in-gauge}_{\mu\nu},
\end{eqnarray}
and the second term results from the integration in the $t'$ variable of the retarded Green function
\begin{eqnarray}
G_{{\rm R}\mu\nu}^{~~\mu'\nu'}(x,y)= \delta_{\mu}^{\mu'} \delta_{\nu}^{\nu'}{1 \over |\mathbf{x}-\mathbf{y}|}\delta(t'-t+|\mathbf{x}-\mathbf{y}|/c),
\end{eqnarray}
where $x=\{t, \mathbf{x}\}$, $y=\{t, \mathbf{y}\}$. Again, we have selected a precise form for the (linear gravity) retarded Green function and used Cartesian coordinates in the selected Minkowski spacetime for clarity. 

Now, for the VBL result to hold, the entire $h^{\rm in}_{\mu\nu}$ must be zero. Among the free solutions of the D'Alembertian equation~(\ref{Eq:LEEs}) there are some $h^{\rm in-phys}_{\mu\nu}$ that represent real gravitational waves (for instance, the $\eta_{\mu\nu} + h^{\rm in-phys}_{\mu\nu}$ metric leads through the Levi-Civita connection to non-zero curvature) while other solutions represent just pure gauge waves associated to the residual gauge symmetry, which in a GR language would correspond to wavy changes of coordinates. 
Indeed, the term $h^{\rm in-gauge}_{\mu\nu}$ represents the residual gauge freedom, namely all transformations that preserve the de Donder gauge condition~\eqref{Eq:DonderGauge} on $h_{\mu\nu}$. It is easy to check that these are transformations of the form: 
\begin{align}
    h_{\mu\nu} \rightarrow h'_{\mu\nu} = h_{\mu\nu} + 2 \nabla_{(\mu} \xi_{\nu)}
\end{align}
where the generator $\xi^\mu$ fulfills the condition
\begin{align}
    \square \xi^\nu = 0. 
\end{align}
Therefore, when the VBL paper states that they have ``tacitly assumed that there is no incoming gravitational radiation'', this is in fact more subtle than just eliminating physically real incoming waves; one has to eliminate also any ``gauge wave''. Otherwise, non-trivial $h^{\rm in-gauge}_{\mu\nu}$ terms could cause the causal cones of $g_{\mu\nu}$ to lie outside the selected Minkowskian reference cones (the cones of $\eta_{\mu\nu}$), even if all matter satisfies the BNEC condition.

In the traditional, geometric GR language we know that any $h^{\rm in-gauge}_{\mu\nu}$ can be absorbed into the arbitrariness of the partition $\eta_{\mu\nu} + h_{\mu\nu}$, since $\eta_{\mu\nu} + h^{\rm in-gauge}_{\mu\nu}$ is in general equivalent to another Minkowskian $\tilde{\eta}_{\mu\nu}$ metric written in suitable coordinates. Therefore, to fix the term $h^{\rm in-gauge}_{\mu\nu}$ to zero is tantamount to restricting completely and uniquely what one takes to be $\eta_{\mu\nu}$. 

Note at this point that, although not mentioned anywhere in the VBL paper itself, the VBL result actually finds a better accommodation within a non-geometric paradigm. We will discuss this issue later on.

\section{The Gao-Wald counterargument}
\label{Sec:GW}

In \cite{GaoWald2000}, Gao and Wald argue that, given a metric $g_{\mu\nu}$ and a decomposition $g_{\mu\nu}= \eta_{\mu\nu}+h_{\mu\nu}$, whether the light cones of $g$ lie inside or outside those of $\eta$ is just a gauge-dependent issue and thus contains no physical information.
First of all, it is crucial to stress that the Gao-Wald analysis assumes the GP explicitly (or equivalently the FTP): They maintain themselves firmly based in general relativity as a gauge theory of a single universal metric. Let us discuss the Gao-Wald position in some detail. 

When standard GR is written as if it depended on two metrics $\eta$ and $g$, the theory acquires additional gauge symmetries besides the standard diffeomorphism invariance. This is easily seen by looking at the following simple example with just two particles (no fields). Consider a Lagrangian of the form
\begin{eqnarray}
L={1 \over 2} \dot{x}^2_1 + {1 \over 2} \dot{x}^2_2 + \dot{x}_1 \dot{x}_2 -V(x_1+x_2).
\end{eqnarray}
This Lagrangian is invariant under the transformation $\delta x_1 = \alpha(t)$ and $\delta x_2 = - \alpha(t)$ with $\alpha(t)$ an arbitrary function of $t$. This system has therefore one gauge symmetry.

Even though the system might appear as having 2 degrees of freedom, $x_1$ and $x_2$, it actually has only one degree of freedom. This can be verified trivially by defining $q=x_1+x_2$ and $r=x_1-x_2$, which allows to rewrite the Lagrangian as
\begin{eqnarray}
L={1 \over 2} \dot{q}^2 -V(q).
\end{eqnarray}
The elimination of a redundant coordinate has automatically made the gauge symmetry $\alpha(t)$ disappear.

Notice the difference with a Lagrangian of the form
\begin{eqnarray}
L={1 \over 2} \dot{x}^2_1 + {1 \over 2} \dot{x}^2_2 -V(x_1+x_2).
\end{eqnarray}
This is invariant only under $\delta x_1 = \alpha$ and $\delta x_2 = - \alpha$ with $\alpha$ constant, not an arbitrary function $\alpha(t)$. Thus, this symmetry is physical, not gauge. 

Within standard GR, the decomposition of the metric $g$ into a background $\eta$ and a perturbation $h$ follows the same pattern just explained. Apart from the usual gauge symmetry (diffeomorphism invariance) of the theory written with just one $g$, there is an additional gauge symmetry associated with the (fictitious) partition into $\eta$ and $h$, which thus could be called a ``background gauge symmetry'' to distinguish it from the former.

Gao-Wald argue that, given a metric $g_{\mu\nu}$ and a decomposition $g_{\mu\nu}= \eta_{\mu\nu}+h_{\mu\nu}$, whether the causal cones of $g$ lie inside or outside those of $\eta$ is just a gauge issue and thus not physical. We can easily see their point by taking the Schwarzschild metric external to a non-compact star of radius $R\gg 2m$ (this is to ensure that we are within a perturbative regime with respect to a flat metric). Written in standard  Kerr-Schild coordinates $\{t,r,\theta,\varphi\}$, the Schwarzschild metric $g_S$ reads
\begin{eqnarray}
ds^2 = \eta_{\mu\nu} dx^\mu dx^\nu+ l_\mu l_\nu dx^\mu dx^\nu;~~~~~~l_\mu=\left\{ -\sqrt{2m \over r} , -\sqrt{2m \over r},0,0 \right\},
\label{kerr-schild}
\end{eqnarray}
thus with $h_{\mu\nu} = l_\mu l_ \nu$ and with the $\eta$ metric written in polar coordinates
\begin{eqnarray}
\eta_{\mu\nu} dx^\mu dx^\nu= -dt^2 +dr^2 + r^2 d\Omega_2^2.
\end{eqnarray}
This setup satisfies the VBL idea that the lightcones of $g_S$ lie inside those of $\eta$: $[g_s] \leq [\eta]$.

Let us add and subtract from the previous expression
\begin{eqnarray}
\tilde{l}_\mu \tilde{l}_\nu dx^\mu dx^\nu+\tilde{k}_\mu \tilde{k}_\nu dx^\mu dx^\nu,
\label{k-l-tilde}
\end{eqnarray}
with
\begin{eqnarray}
\tilde{l}_\mu=\left\{ -C,-C,0,0 \right\},~~~~~
\tilde{k}_\mu=\left\{ -C,C,0,0 \right\},
\label{k-l-tilde-II}
\end{eqnarray}
where $C$ is a constant such that $C^2 > 2m/R$.
One can easily check that the metric 
\begin{eqnarray}
\tilde{\eta}_{\mu\nu}  = 
\eta_{\mu\nu}  + \tilde{l}_\mu \tilde{l}_\nu  + \tilde{k}_\mu \tilde{k}_\nu 
\label{eta-tilde}
\end{eqnarray}
is again a flat metric\footnote{Strictly speaking, this metric has a cone singularity in $r=0$, but this is irrelevant for the illustrative purpose of our example.}, but we have managed to close inwards (in the radial direction) its lightcones with respect to the former flat metric, $[\tilde{\eta}] \leq [\eta]$, as can easily be seen by checking that $\tilde{\eta}_{\mu\nu} l^\mu l^\nu  \geq 0$ for any null vector with respect to $\eta$. 

Of course, $g_S$ has not changed under the operation \eqref{k-l-tilde}. The assignation \eqref{eta-tilde} is compensated by the new $h_{\mu\nu}$ piece, that we will denote by $\tilde{h}_{\mu\nu}$, which reads
\begin{eqnarray}
\tilde{h}_{\mu\nu} = h_{\mu\nu} - \tilde{l}_\mu \tilde{l}_\nu - \tilde{k}_\mu \tilde{k}_\nu.
\end{eqnarray}
This piece contributes to open up the cones of $g_S$ with respect to $\tilde{\eta}_{\mu\nu}$. The constant $C$ has been set so that now $ [g_S] \geq [\tilde{\eta}]$, and thus, with respect to the new flat referential metric  $\tilde{\eta}$, the metric $g_S$ has its causal cones outside.

Which part of the VBL theorem has precisely been violated in this manipulation? Strictly speaking the former $h_{\mu\nu}$ is already non-harmonic, but the really crucial element is that in the change from $h_{\mu\nu}$ to $\tilde{h}_{\mu\nu}$, we have introduced a non-trivial $h^{\rm in-gauge}_{\mu\nu}$.

The example provided by Gao-Wald is based on a linear gauge transformation \mbox{$\delta h_{\mu\nu}=2\nabla_{(\mu} \xi_{\nu)}$}. 
There is a subtle point in performing this gauge transformation that we will now explain.
Notice that at the non-linear level the gauge symmetry of GR becomes identified with diffeomorphisms or coordinate transformations. If we apply a non-infinitesimal coordinate transformation to $\eta_{\mu\nu} + h_{\mu\nu} $ we have  
\begin{eqnarray}
\eta'_{\mu'\nu'}(x') + h'_{\mu'\nu'}(x')= (\eta_{\mu\nu}(x) + h_{\mu\nu}(x)) T^\mu_{\mu'}T^\nu_{\nu'},
\end{eqnarray}
with 
\begin{eqnarray}
T^\mu_{\mu'}= {\partial x^\mu \over \partial x'^{\mu'} }.
\end{eqnarray}
The matrix transformation applies to both  $\eta_{\mu\nu}$ and $h_{\mu\nu}$. At the infinitesimal level, the transformation of coordinates can be written as
\mbox{$x'^\mu =x^\mu + \xi^\mu (x)$}, and we have 
\begin{eqnarray}
\eta'_{\mu'\nu'}(x) + h'_{\mu'\nu'}(x) = \eta_{\mu\nu}(x) + I^{(1)}_{\mu\nu}(\xi) + h_{\mu\nu}(x) + I^{(2)}_{\mu\nu}(\xi).
\end{eqnarray}
It is well known that
\begin{eqnarray}
I^{(1)}_{\mu\nu}(\xi) + I^{(2)}_{\mu\nu}(\xi)= 2\nabla^g_{(\mu} \xi_{\nu)}.
\end{eqnarray}
Gao-Wald decided to use the presence of the additional background gauge symmetry to apply the entire transformation just to $h_{\mu\nu}$.
To be even more clear, we can perform first a standard (gauge) coordinate transformation
\begin{eqnarray}
&&\eta_{\mu\nu} \to \eta_{\mu\nu} + I^{(1)}_{\mu\nu}(\xi),  
\\
&&h_{\mu\nu} \to h_{\mu\nu}+ I^{(2)}_{\mu\nu}(\xi),
\end{eqnarray}
and then a background gauge transformation
\begin{eqnarray}
&&\eta_{\mu\nu} \to \eta_{\mu\nu} - I^{(1)}_{\mu\nu}(\xi),  
\\
&&h_{\mu\nu} \to h_{\mu\nu}+ I^{(1)}_{\mu\nu}(\xi).
\end{eqnarray}
In this way we have the following total transformations
\begin{eqnarray}
&&\eta_{\mu\nu} \to \eta_{\mu\nu} 
\\
&&h_{\mu\nu} \to h_{\mu\nu}+  I^{(2)}_{\mu\nu}(\xi) + I^{(1)}_{\mu\nu}(\xi) = h_{\mu\nu}+2\nabla^g_{(\mu} \xi_{\nu)},
\end{eqnarray}
which at the linear level become 
\begin{eqnarray}
&&\eta_{\mu\nu} \to \eta_{\mu\nu} 
\\
&&h_{\mu\nu} \to h_{\mu\nu}+2\nabla_{(\mu} \xi_{\nu)}.
\end{eqnarray}
This is precisely the transformation scheme used by Gao-Wald. If the initial $h_{\mu\nu}$ satisfies the condition 
\begin{eqnarray}
\nabla_\mu (h^{\mu\nu}-{1 \over 2}h\eta^{\mu\nu})=0
\end{eqnarray}
then the new $h_{\mu\nu}$ will satisfy
\begin{eqnarray}
\nabla_\mu (h^{\mu\nu}-{1 \over 2}h\eta^{\mu\nu})=\square \xi^\nu,
\end{eqnarray}
which in general will not be zero. Gao-Wald present an example of a transformation of the form 
\begin{eqnarray}
\xi_\nu = -{r_\mu \over 2 + f(r)},
\end{eqnarray}
defined in polar coordinates in a Minkowxki spacetime. Here $r^\mu$ is the radial four-vector and $f(r)$ is a function such that that $f(r)= r^2$ for $0 \leq r  \leq 1/2$, $f(r)= r$ for $r \geq 1$, and in between any smooth interpolation with 
$f(r) \geq 0 $ and $0 \leq f'(r)< 2$. For instance, they show that this gauge transformation opens up everywhere the lightcones associated with $g$ with respect to those associated with $\eta = g-h $. They do not care whether this transformation maintains or not the harmonic gauge (in fact, they recall a result~\cite{GerochXanthopoulos1978} showing that the harmonic condition is ill-defined at null infinity; a point to which we will come back later). But even if one finds a gauge transformation that preserves the harmonic gauge, the problem of the pure residual gauge additions described above will still persist.

Therefore, to summarize the central message of Gao-Wald: to be or not to be inside the causal cones of a flat reference metric is a highly gauge-dependent notion and thus not appropriate to construct physical connections, for instance, between superluminal behaviour and EC violations.

\section{The harmonic background paradigm}
\label{Sec:ThirdP}

The Gao-Wald argument is absolutely correct within the geometric and field-theoretic paradigms. If one maintains this view, any attempt to establish a hierarchy between the causal cones of a general relativistic metric $g$ and a Minkowski metric $\eta$ extracted from $g$ is doomed to fail.
But is there a way to make the VBL vision consistent? We think there is. The key point, in our opinion, lies in introducing bi-metricity and taking its concept and implications further than in the standard field-theoretic paradigm.

In a field-theoretic paradigm one assumes that there exists a flat metric on top of which a tensor field $h_{\mu\nu}$ is evolving.
The dynamical equations of the tensor field contain non-linearities. More importantly, they also exhibit a gauge symmetry.
This gauge symmetry is precisely 
\begin{eqnarray}
\delta h_{\mu\nu} = 2\nabla^{\eta+h}_{(\mu} \xi_{\nu)}.
\end{eqnarray}
Note that, from the point of view of other gauge symmetries in field theories, such as Yang Mills fields, it can be argued that this gauge symmetry is a bit odd as it depends on the very $h$ field itself. But the key point is that in this paradigm there is still no way to attribute a physical reality to the relation between $\eta$-cones and $(\eta+h)$-cones, precisely because it is a gauge-dependent notion. There is another important notion that is not gauge-invariant, and therefore lacks physical meaning: to associate to a curve in the $(\eta+h)$ space a single curve as seen from the point of view of $\eta$. Indeed, take for instance a geodesic of $(\eta+h)$. 
Changing $h_{\mu\nu}$ by a gauge transformation changes the form of the geodesic equation 
%
\begin{eqnarray}
{d^2 x^\sigma \over d\tau^2} + \Gamma_{\mu\nu}^\sigma(h) {d x^\mu \over d\tau}{d x^\nu \over d\tau}=0.
\end{eqnarray}
Therefore, given a specific geodesic trajectory of $(\eta+h)$, it cannot be associated with a unique curve from the point of view of hypothetical observers who only perceive the $\eta$ metric. The interpretation within the GP is that the trajectory of a celestial body $x(\lambda)$ makes sense only in its relation with other trajectories, but not ``in itself''. In other words, there is no unique way to construct a referential system. 
In the context of perturbation theory in GR it is well known that the selection of a gauge is tantamount to the identification of the points of two metric manifolds, the physical metric and the background metric~\cite{Brunietal2002}.

A related question is that of the stress-energy tensor (SET) for the gravitational field $h$.
Already Rosen~\cite{Rosen1940a} realized that adding an extra flat structure allows to construct a proper SET for $h$ as opposed to a pseudo-tensor. 
However, he did not mention in this paper that this SET is not gauge-invariant, not even at the linear level. In GR this is easily interpreted as the absence of a local notion of gravitational energy. Thus, this is another instance of notions that one would like to have when building a field-theoretic view, but these notions fail to acquire physical meaning because they are not gauge invariant. In this sense, the geometric and the field-theoretic paradigms are closer to each other than one could expect at first glance. From a conceptual point of view there is not much advantage in using the field-theoretic paradigm. In fact, the geometric paradigm provides a simpler interpretative framework to understand such conundrums as the cone hierarchy, relational trajectories and non-local energy.

However, one can solve all these quandaries by taking the field-theoretic paradigm one step further. One direct manner to give complete physical reality to the $\eta$ background metric is to introduce a physical or privileged gauge for $h_{\mu\nu}$. Or, from a different perspective, to assume that we do not have a gauge theory of gravitation but a theory of a tensor field $h_{\mu\nu}$ subject to certain restrictions. Before describing our specific proposal, which the reader can foresee based on our discussion so far, let us make two digressions, one on the notion of diffeomorphism invariance in the context of bi-metrical theories, and the other on Fock's harmonic coordinates.

\subsection{Diffeomorphisms and bi-metricity}
\label{Subsec:Diff}

Let us a recall a few basic notions. A pseudo-Riemannian manifold is a manifold endowed with a metric tensor of Lorentzian signature. Two pseudo-Riemannian manifolds, $\{ {\cal M}, g \}$ and $\{ {\cal M}',g' \}$, are diffeomorphic 
\begin{equation}
\{ {\cal M}, g \} \sim \{ {\cal M}', g' \}
\end{equation}
if ${\cal M}$ and ${\cal M}'$ have the same topology and there exist coordinates $x$ in ${\cal M}$ and $x'$ in ${\cal M}'$, which might be defined only patch by patch, such that $g(x)$ has the same functional form as $g'(x')$. Note that we say ``diffeomorphic'': $\{ {\cal M}, g \} \sim \{ {\cal M}', g' \}$, not ``equal'': $g=g'$. 

In mathematical terms, two different metrics in a single manifold represent in principle two different structures, even if they were diffeomorphic when considered in two manifolds. However, to this initial mathematical structure we can add additional features motivated by the physics one wants to represent. For instance, if one considers that the physics does not offer any operational way to distinguish two diffeomorphic metrics in a single manifold, for example assuming a diffeomorphism-invariant dynamical theory for the metric, then one can take one step further and consider that the real physical objects are not the metrics themselves but equivalence classes of metrics related by diffeomorphisms: 
\begin{equation}
\{ {\cal M}, [[g]] \}
\end{equation}
where the double brackets represent the idea of an equivalence class. Then, we can say that instead of a \emph{metrical theory} we have a \emph{geometrical theory}. This is the usual physical interpretation of General Relativity, which of course goes back to Einstein himself: only diffeomorphism invariant properties have physical reality. 
A consequence of this point of view, as mentioned above, is that it only allows to define relational quantities. This can be taken as a fundamental characteristic, but also as a serious drawback. For instance, a concept such as gravitational energy is in principle ill-defined, since it cannot be described in any straightforward way as a tensor.
Moreover, GR has no local observables: all the physical (i.e.\ gauge-invariant) quantities are highly non-local~\cite{Bergmann1961gauge}. This relational interpretation is in stark contrast with the way in which many problems in GR are analyzed in practice, such as the geodesics of bodies around a black hole: to understand these problems one typically fixes a particular set of coordinates.  

Consider now a manifold with two pseudo-Riemannian metrics $\{ {\cal M},g_1,g_2 \}$. We can then follow the same steps as in the 1-metric case, up to a certain point. We can define a notion of diffeomorphic equivalence between $\{ {\cal M}, g_1, g_2 \}$ and another structure $\{ {\cal M}',g'_1,g'_2 \}$:
\begin{equation}
\{ {\cal M},g_1,g_2 \} \sim \{ {\cal M}',g'_1,g'_2 \}
\end{equation}
under the adequate conditions, which so far are a relatively straightforward extension from the 1-metric case. We can consider that physical theories are not able to discern diffeomorphisms. We can then use the freedom in the diffeomorphism group to consider, for example, $g_1$ as a geometry instead of as a metric, thus worrying only about equivalence classes of $g_1$: 
\begin{equation}
\{ {\cal M},[[g_1]],g_2 \}
\end{equation}
However, at this point, it is not straightforward at all to consider that one should also necessarily deal with geometric equivalence classes $[[g_2]]$. In fact, it is more natural that $g_2$ will continue to operate physically as a metric. Only in very specific cases, for example when the dynamical theory has a second, independent group of symmetries, assimilable to diffeomorphism invariance, could one speak of a genuinely ``bi-geometric'' theory. 
In generic dynamical theories with two metrics and differmorphism invariance, once one assumes that $g_1$ is a geometry, the second metric $g_2$ in principle maintains more physical degrees of freedom than a geometry. 
However, we know that in Nature there is a regime well approximated by the GR equations where no additional (i.e.\ metrical) degrees of freedom appear. We also know that these GR equations find an excellent accommodation in both the geometric and the field-theoretic paradigms. 
It thus seems obvious to require that $g_2$ should at some point stop being a completely arbitrary metric tensor to start looking more like a geometry. This is obtained by imposing four  restrictions on $g_2$ (or $h$, in our construction), which mimic the elimination of the freedom associated with changes of coordinates in 4 dimensions. This requirement is by no means sufficient to obtain GR, nor is it compulsory (for example, one could have a strict bi-geometrical theory in which at some point one metric becomes operationally invisible), but we think it is interesting to appreciate its naturalness.

\subsection{Fock harmonic condition}
\label{Subsec:Fock}

Our investigation into these topics started by looking at the VBL/Gao-Wald controversy. However, in revising the scientific literature we encountered several ideas which resonated with our own proposal. One of them deserves special mention because of its relevance and strong connection with our own proposal. This is the notion of harmonic coordinate system put forward by Fock~\cite{Fock1959}. This section is presented from the point of view of the geometric or field-theoretical paradigms.

Consider a Minkowski spacetime and some Cartesian coordinates $x^\mu$. Let us solve the D'Alembertian equation $\square_\eta X^{(\mu)}(x)=0$ for 4 scalar functions $X^{(\mu)}=X^{(\mu)}(x)$. The general solution of this equation is
\begin{eqnarray}
X^{(\mu)} = C^{(\mu)} + \Lambda^{(\mu)}_{~\nu} x^\nu + \int d\mathbf{k} \left[A_{\mathbf k}^{(\mu)} \cos(wt-{\mathbf k} \cdot {\mathbf x})
+ B_{\mathbf k}^{(\mu)} \sin(wt-{\mathbf k} \cdot {\mathbf x}) \right],
\end{eqnarray}
with $C^{(\mu)},\Lambda^{(\mu)}_{~\nu},A_{\mathbf k}^{(\mu)},B_{\mathbf k}^{(\mu)}$ different constants. If we remove any wavy solutions and set the trivial constant term to zero (corresponding to a shift in the origin of coordinates) we are left with some special solutions   
\begin{eqnarray}
X^{(\mu)} = \Lambda^{(\mu)}_{~\nu} x^\nu 
\end{eqnarray}
which are linear combinations of the Cartesian coordinates themselves. In more detail, Fock proved the following:\\
Given a global Cartesian system of coordinates, any other harmonic set of coordinates, i.e.\ $\square_\eta X^{(\mu)}(x)=0$, which
\begin{enumerate}[label=\roman*)]
    \item is asymptotically Cartesian, i.e.\ any departure from Cartesianity dies off as \mbox{$f(x) \sim 1/r$}, where $f(x)$ represents any of the coordinate functions,
    \item is bounded, i.e.\ any departure from Cartesianity obeys $f(x)<C$, and
    \item contains at most outgoing waves, i.e.
\end{enumerate}
\begin{eqnarray}
\lim_{r \to \infty} \left\{ 
{\partial (r f) \over \partial r} + {1\over c}{\partial (r f) \over \partial t}
\right\}=0,
\end{eqnarray}
then this new set of coordinates is related to the original set of Cartesian coordinates by a global Lorentz transformation. For this reason, we can say that Cartesian coordinates are precisely the unique non-wavy solutions of the harmonic equation, or that Cartesian coordinates are harmonic for the flat geometry. 

From a dual perspective, instead of imposing restrictions on the coordinates, one could impose restrictions on the form in which the metric $\eta$ is expressed. If we write $\eta$ in (in principle: arbitrary) coordinates $X^{(\mu)}$ and impose the restriction that these coordinates are harmonic: 
\begin{equation}\square_\eta X^{(\mu)}=0,\end{equation}
then, this restriction amounts to the condition $\eta^{\mu\nu} \Gamma^{\sigma}_{\mu\nu}=0$ or equivalently $\partial_\mu \sqrt{-\eta}\eta^{\mu\sigma}=0$. Thus, the harmonic condition restricts the forms that the flat metric can acquire. To single out a unique form for the metric, one has to add additional restrictions on $\eta$ besides the harmonic condition: essentially that it does not contain wavy components.    

Fock then argues that the same uniqueness should hold true for asymptotically flat spacetimes under equivalent conditions. We can see that that is true at least for weakly curved spacetime, that is, metrics of the form $g=\eta+h$. The equation $\square_g X^\mu(x)=0$ can be written also as
\begin{eqnarray}
\square_\eta X^\sigma(x)+ g^{\mu\nu} C^\alpha_{\mu\nu} \partial_\alpha X^\sigma =0,
\end{eqnarray}
where $C^\alpha_{\mu\nu}$ is a tensor which has the same structure of the Levi-Civita symbols but with $\partial_\mu$ substituted by $\nabla_\mu$. If we perform the calculation with coordinates $X^{(\mu)}$ which are harmonic with respect to the flat metric, then from the previous expression we obtain the consistency condition \begin{equation}g^{\mu\nu} C^\alpha_{\mu\nu}=0.\end{equation} But at first order this corresponds precisely to the harmonic gauge condition (\ref{Eq:DonderGauge}).
The additional condition to prescribe a unique system of coordinates (modulo global Lorentz transformations) is to remove from $h_{\mu\nu}$ any incoming wave solution, which in the linear case correspond to free waves not originating from any source.

So Fock's idea, to which we here completely subscribe, can be summarized as follows: at least for weakly curved metrics, one can select a unique and privileged set of coordinates, which Fock called the {\em harmonic coordinate system}. Obviously, one can still perform calculations in any other coordinate system. But Fock strongly emphasized the significance of the existence of this privileged system of coordinates in terms of extracting physical information from gravitational systems. For example, if a unique privileged system of coordinates still exists in the non-linear regime, then it can be used to define precisely what it means to be in a linear regime, namely that the components of $h_{\mu\nu}$ must be small with respect to the unit prescribed by the value of the components of the $\eta$ metric.



\subsection{The paradigm}
\label{Subsec:Paradigm}

We are now ready to present what we call the ``harmonic background paradigm''. The paradigm is constituted by a a series of hypotheses, or fundamental assumptions. The idea would then be to develop the consequences of these hypotheses and to check their predictive power. 

The first assumption is that there exists something playing the role of a flat background geometry, represented by a metric $\eta_{\mu\nu}$ written in whatever system of coordinates. With this assumption, we are not committing to an ontological reality of this spacetime, only that such a structure exists pragmatically as a way of describing the fundamental causality in the system. 
We think of $\eta$ as a fixed background geometry, not in an absolutely rigid sense, but in the sense of an effective decoupling because of the large ``inertia'' of this background with respect to all the other fields. The background metric $\eta$ influences the other fields (``geometry tells effective fields how to move''), but these fields barely affect the background metric (``field content tells the geometry how to curve'' is no longer valid for the background metric $\eta$). Notice that situations of this sort are very standard in the physical description of phenomena. Physically speaking, we can think of this background flat geometry as a constant extrapolation to the entire manifold of the geometry in regions very far away from any source and in moments in which there are no gravitational waves passing by. Once we have this background geometry installed, we will impose that no effective causality can surpass its propagation speed limit.

Our second assumption is that gravity is represented by a tensor object $h_{\mu\nu}$ living on this geometrically flat manifold. In many circumstances in which $h_{\mu\nu}$ is sufficiently small compared to $\eta_{\mu\nu}$, we have that $g_{\mu\nu}:=\eta_{\mu\nu} + h_{\mu\nu}$ can be considered formally as a second pseudo-Riemannian metric. If $h_{\mu\nu}$ is too big, $g_{\mu\nu}$ might not even be a Lorentzian metric. Note that this regime need not coincide with the linear regime. Now, for the reasons explained above, we do not want to consider this second metric as a second \emph{geometry} but something very close to that notion. Instead of working with a completely general tensor $h_{\mu\nu}$, we will therefore from the very start impose several restrictions on 
$h_{\mu\nu}$. On the one hand, we impose four differential conditions. At the linear level, these conditions are precisely the de Donder conditions (\ref{Eq:DonderGauge}). We leave their extension to the full non-linear regime for future work. This extension need not be in the standard non-linear harmonic condition. Note that in fact we do not need the non-linear regime to find one of the principal results of this paper.

Finally, as an additional restriction we consider that the physical solutions of $h_{\mu\nu}$ do not contain waves of any kind coming in from infinity, neither physical nor gauge waves. Note that this additional condition on the allowed forms of $h_{\mu\nu}$ cannot be imposed through a local differential condition but requires a boundary condition. These restrictions on $h$ plus 
the presence of a background are precisely what inspire the name of the paradigm.

Therefore, the paradigm we propose here is essentially bi-metric; it assigns an ``ontological'' priority to the $\eta$-geometry with respect to the gravitational $g$ metric; it considers that gravity is not described by a gauge theory but instead assumes that there is a unique $h_{\mu\nu}$ representing each gravitational situation. Thus, we take Fock's idea very seriously (although Fock's own philosophical view on these matters is not completely clear to us): beyond considering that there exists a privileged gauge, our language assumes a specific definition of what is the gravitational field. One can not impose any gauge one likes, in fact, in our view such an assertion makes little sense: we take as hypothesis that the restrictions on $h$ are physical, not originating from any gauge fixing. 
For example, if the unique $h$ becomes singular, then this singularity should be considered physically relevant; it would make no sense to verify ``whether the singularity disappears in a different gauge'' because the theory is not a gauge theory. 

Of course, whenever the phenomenology remains in a weak-field GR regime, the three paradigms we have been discussing (geometric, field-theoretic, and harmonic background) can be considered as just three interchangeable and equivalent approaches, with some calculations easier in one or another paradigm. So, from the perspective of the harmonic background paradigm, most calculations performed within a gauge perspective make perfectly good sense; only the interpretation would be different. However, as we will see in the next section, even in standard situations there can be important differences of interpretation, in particular the harmonic background paradigm can be understood as pointing towards an explanation of why gravity is attractive. 
Nonetheless, the more crucial distinction between these paradigms really crystallizes when one tries to go beyond the GR regime, for instance, when trying to solve the singularity problem.
The harmonic background paradigm gives an important role to the background geometry and thus suggests that the first departures from GR might appear precisely because of this prominence of the background geometry (for instance, in terms of effective models and situations not describable in terms of just one single $g=\eta+h$).
For example, it is well known that trying to incorporate even a tiny mass for the graviton forces one introduce an additional metric besides the GR metric~\cite{deRham2014}. In a bi-metric paradigm, GR can be seen as a small-mass limit of contemporary massive-gravity dynamical theories. From this point of view, a bi-metrical formulation of GR appears as more robust and less isolated in theory space.

Now, by this point, it should be obvious that, by construction, the VBL result is absolutely and rigorously valid in the harmonic background paradigm. Moreover, this paradigm not only allows to establish a hierarchy of causal cones whenever the Background NEC is satisfied (at least at the linear level). It also allows to define a unique SET for all the physical gravity tensors $h_{\mu\nu}$. Furthermore, it permits to associate a single trajectory in the flat background to the geodesic movement of any celestial body. All these highly desirable characteristics are completely absent in the other two paradigms in which gauge symmetries play a crucial role. 

Let us also stress that the harmonic background paradigm is compatible with situations in which there is emission of gravitational waves (these waves are outgoing, not incoming). The full conceptual consistency of the paradigm occurs when considering bounded and isolated closed systems. If one wants to analyze in simple mathematical terms what happens when a gravitational wave generated far away interacts with some local system, it might be convenient to relax the formalism and allow the presence of physical waves incoming from infinity. Here we just want to remark that this calculational convenience should not be confused with a physical fact. 

To end this section and re-connect with the VBL argument, let us recall that the VBL paper was originally formulated within the geometric paradigm, and correctly critized within that same paradigm by Gao-Wald. Note that the VBL argument can also be understood within the field-theoretic paradigm that we mentioned earlier. But even in such a field-theoretic view, it does not acquire a convincing power, since it would still be gauge-dependent. The alternative paradigm that we have discussed here stresses the importance of the flat background geometry and the specifically restricted $h$. It is thus within this paradigm that the VBL idea acquires a renewed rigour and interest.

\section{Why is gravity attractive?}
\label{Sec:Why-attractive}

We started this paper with the interesting debate between VBL and Gao-Wald, explaining why both are correct. The clash appeared because of their different initial positions. 
While Gao-Wald maintained themselves clearly and completely within the geometric paradigm, or equivalently the field-theoretic paradigm (but in any case a gauge-independent formalism), 
the VBL paper can be considered as a cautious but insightful first heuristic step in the exploration of extensions of GR beyond these paradigms. We have shown that the VBL proposal can be made totally rigorous in a harmonic background setup. In this framework gravity appears as a modification of a more rigid background causality described by a flat background Minkowski metric. 

In the present section, we will discuss a result that immediately follows from the harmonic background paradigm simply by inverting the logical arrow of the VBL paper. VBL emphasized that the satisfaction of an energy condition forbids superluminal propagation. Here we explain, vice versa, how the existence of a fundamental background causality can be interpreted as automatically imposing an energy condition.

GR is a peculiar theory. It establishes a connection between matter/energy content and spacetime curvature, but it says nothing about what kind of matter can exist in Nature. In fact, the distinctive attractive character of gravity is not explained by GR, nor by Newtonian gravity: it is an additional postulate about the positive nature of matter/energy.
Curiously, though, the proof of many of the most important results in GR requires this additional postulate, i.e.\ the satisfaction of some energy condition. 

The characteristics of the harmonic background paradigm as described before suggest to add one additional assumption to its very definition. Given the crucial role we are assigning to the flat background geometry, it is most reasonable to impose as a restriction that the effective causality is always subsumed into a more fundamental and insurmountable background causality, in other words 
\begin{equation}[g] \leq [\eta],\end{equation}
always.
When $h$ is small, we are in fact probing the possible first deviations from a flat causality. It is an empirical fact that, in all usual low-gravity situations, mass and energy are always positive. The VBL result tells us that the BNEC, or background NEC, can be interpreted (at the linear level) as a consistency condition for the absence of superluminality, i.e.\ $[g] \leq [\eta]$. We have discussed in detail that the VBL result automatically follows from the harmonic background paradigm. We therefore propose to interpret $[g] \leq [\eta]$, i.e.\ the presence of a more fundamental background causality, as the real reason why gravity is attractive, at least in normal weak-gravity situations. The idea is that the obstruction to make the $g$-cones exit the $\eta$-cones restricts the stress-energy characteristics that any matter content may have.

Note however that this energy condition need not necessarily be maintained in regions where the cones of $g$ are already well inside those of $\eta$, in other words in the non-linear regime for $h$. 
First, from a technical point of view, it appears that neither the VBL result nor the harmonic condition can be extended in any straightforward way to the non-linear case. But beyond this technical issue, there is a more profound reason why energy conditions might be violated in high-curvature regimes. There is nothing in the harmonic background paradigm which indicates that energy conditions should always be imposed. Within this paradigm, it is only when the effective cones approach the Minkowski cones that one should start appreciating an obstruction or constraint to the behaviour of matter. That is, precisely in the weak-field regime to which our analyses are restricted.

Beyond this regime, although not straightforward to prove, we suspect that imposing suitable energy conditions under all circumstances would indeed force the $g$-cones to become narrower and narrower. In such a picture, the formation of a singularity seems inevitable in situations such as a sufficiently massive standard gravitational collapse. Our framework, on the contrary, suggests to appeal to the background metric, and to proper bi-metrical generalizations of the GR regime, whenever the effective field $h_{\mu\nu}$ approaches a singular situation. Such ideas and how they could help to avoid the formation of singularities in gravitational collapse have been explored in \cite{Barceloetal2011,Barceloetal2015,Barceloetal2016}. There we advocate for the presence of a flat beckground geometry
and to characterize the strong gravity regime as a switching off of the effective causality, leading for example to black-hole to white-hole bounces. 

To summarize this section, let it suffice to repeat that the harmonic background paradigm provides an explanation, which in our opinion is very natural, for the attractive character of gravity, as an immediate consequence of the fundamental causality of the background structure.

\section{About the Penrose property}
\label{Sec:Penrose}

We cannot finish this paper without mentioning and briefly discussing a potential problem with the harmonic background paradigm and in particular the assumption that it contains a cone hierarchy. The problem comes from Penrose's argument that GR as a geometric theory cannot be reduced simply to a Lorentz-covariant field theory~\cite{Penrose1980}. The essence of the problem has to do with the possibility or impossibility of interpreting gravity as a dynamical theory of cones always inside some fixed Minkowski structure, $[g]\leq [\eta]$, a point with an obvious relevance with respect to our previous discussion.  
We will explain in this section why Penrose's idea does not pose an essential problem for our proposal.

Let us start by reviewing the logic of Penrose's argument, see also its review and extension in~\cite{CameronDunajski2020}. Penrose implicitly works with a single manifold with two metric structures, $\{ {\cal M},\eta, g \}$, so that changes of coordinates in the manifold do not change the potential hierarchy between the cones of the two metrics. So far, this view is completely in tune with our harmonic background paradigm presented above, see Sec.~\ref{Sec:ThirdP}, and not so much with a field-theoretic view, precisely because gauge transformations can affect the cone hierarchy.     

Second, he defines what the authors of~\cite{CameronDunajski2020} call the {\it Penrose property}, which is obeyed by a spacetime if and only if given two arbitrary timelike curves there is always a third timelike curve that intersects them both. Third, Penrose proves that Minkowski spacetime does not satisfy the Penrose property, contrarily to physically sensible geometries in 3+1 dimensions, for instance, stellar structures with a positive mass and a Schwarzschild exterior. Finally, based on the previous point, he proves that it is impossible to define a conformal compactified manifold and two compactified metrics 
$\{ \hat{\cal M}, \hat{\eta}, \hat{g} \}$ such that $[\hat{g}] \leq [\hat{\eta}]$.

The authors of~\cite{CameronDunajski2020} argue that Penrose's proof also implies the impossibility of imposing a condition $[g] \leq [\eta]$. However, the crucial observation of Penrose is that any spacetime that approaches Minkowski spacetime at infinity in a  $1/r$ Schwarzschild manner makes any radial null ray acquire an infinite delay in reaching infinity with respect to what would have happened when traversing instead a purely Minkowskian region. This can be seen by looking at the radial null geodesics outgoing from any positive central matter:
\begin{eqnarray}
u = t - r -2M \ln \left(  {r -2M \over 2M} \right).
\end{eqnarray}
Consider any null particle (a photon) escaping away from the central mass towards infinity and already very far from the mass. Compare this with a second photon, equally far from the central mass but coming in from an antipodal position. 
Then, the continuously accumulated delay of the first photon means that the second photon will be able to catch up with the first one by following a trajectory that avoids the center of the configuration. The infinite delay accumulated with respect to pure Minkowski is therefore a characteristic of any asymptotically Schwarzschild geometry. However, the infinite character of the involved spacetimes is essential for this to hold.  To be more explicit, take a conformally compactified Minkowski spacetime, which amounts to a 4D diamond with boundary. Then, prescribe a second metric $\hat{g}$ on this compactified manifold. In order for $\hat{g}$ to incorporate delays with respect to $\hat{\eta}$, assume $\hat{g} \leq \hat{\eta}$. But null rays moving according to $\hat{g}$ will never accumulate an infinite delay with respect to the Minkowski rays in a finite amount of space. Such infinite delay could be obtained if cones of the initial Minkowski compactified representative $\hat{\eta}$ completely open up at the conformal boundary. But then this conformal Minkowski metric no longer possesses the appropriate null infinities. Therefore, it is true (as Penrose demonstrates) that $\{ \hat{\cal M}, \hat{\eta}, \hat{g} \}$ with $[\hat{g}] \leq [\hat{\eta}]$ is impossible. However, and contrarily to the claim made in ~\cite{CameronDunajski2020}, $\{ {\cal M}, \eta, g \}$ with $[g] \leq [\eta]$ {\it is} possible.

Penrose's correct analysis should not be taken as proof that there is an inconsistency in treating GR from the point of view of a Lorentz covariant construction. More precisely: Penrose does not prove it impossible to construct a non-compact manifold in which standard physical metrics, such as those associated with regular stars, are inside a Minkowski metric (in fact, his paper starts by showing precisely an example in which this happens). What is not possible is to compare the causal asymptotic structures of Minkowski and of physical, Schwarzschild-type metrics. 

But how relevant is this for actual physical situations? Our viewpoint is that asymptotic structures in GR allow a better mathematical control over the theory. The same happens in Quantum Field Theory: a mathematical control of scattering theory requires the definition of asymptotically free states. But care should be taken when physical conclusions depend crucially on these asymptotic properties. In particle collider experiments, the asymptotically free states do obviously not exist physically. The fact that such theories work can be considered precisely as a demonstration that the existence of a strict asymptotic region does not play any role. 
In the gravitational case, let us stress that the Penrose property strongly depends on the $1/r$ behaviour of the Newtonian potential in (3+1) dimensions (e.g. in~\cite{CameronDunajski2020} it is shown that stellar configurations in higher dimensions do not exhibit the Penrose property). It is also worth recalling that $1/r$ interactions (in other words, massless fields) always have subtleties in scattering theory~(see e.g. \cite{Newton1982}). Thinking of GR as the small-mass limit of a massive gravity theory, for example, could immediately solve this problem.

Let us take the opportunity offered by this discussion to clarify that with our proposal we are not advocating for the reality of a strict Minkowski background (i.e.\ spatially infinite). We take this simple background as a proxy for more realistic spatially compact settings. For instance, we leave for a future work to analyze if and how the harmonic background paradigm described here could be generalized to spatially compact backgrounds such as an Einstein static universe.  

To end this section let us also briefly comment on another potential challenge for the harmonic background paradigm, as briefly mentioned above in Section~\ref{Sec:GW}. In ~\cite{GerochXanthopoulos1978} the authors discuss that the evolution equation for the perturbations over conformally compactified asymptotically simple spacetimes appears ill defined if one uses the harmonic gauge for the perturbations. Here, let us just say that this does not imply that the harmonic condition applied to perturbations of the physical metric is ill-defined.

\section{Conclusions and further remarks}
\label{Sec:Summary}

Our investigation started by trying to understand a result by Visser-Bassett-Liberati~\cite{VisserBassettLiberati2000} and a subsequent criticism by Gao-Wald~\cite{GaoWald2000}. VBL found that under certain conditions the satisfaction of an energy condition forbids propagation faster than light. Gao-Wald counter-argued that this result is highly gauge dependent and therefore has no physical significance. Here, we have cleared up this debate and reached the conclusion that both were right but that a proper defense and a correct interpretation of the VBL result requires a paradigm shift, for which we have proposed a harmonic background paradigm. 

The relativistic revolution of the 20th century has made an anathema of thinking about any kind of fundamental background, anything that acts but cannot be acted upon~\cite{smolin2006case}. \footnote{Note that, contrarily to common lore, this was not a position that Einstein himself strongly defended, see for example our discussion in~\cite{barcelo2008real}.
} Even bi-metrical theories are usually interpreted as being better motivated if both metrics are dynamical~\cite{baccetti2012massive}. However, if one agrees that our current knowledge of spacetime and gravitation is purely effective, then it might be necessary to take one step back when thinking about a more fundamental theory (at the Planck scale, say), let alone any hypothetical ``final theory'' (whatever that may mean). We believe it only very reasonable to hypothesize that such a more fundamental structure might be sufficiently rigid with respect to our effective physics that it can be treated as a background structure. 

Concretely, in this work we propose an alternative way to understand the gravitational phenomena. We propose that  
the gravitational field (at least at the linear level) is a tensor field $h_{\mu\nu}$ living in a flat background geometry. 
In the GR regime this tensor is sourced by the stress-energy tensor of matter fields and acts as an effective and universal deformation of the causal structure of the background geometry: all matter fields couple to $\eta+h$. The tensor $h_{\mu\nu}$ solves the Einstein equations and in addition is subject to three restrictions: i) it satisfies the harmonic condition \mbox{$\nabla^h_{\mu\nu}- (1/2) \nabla_\nu h=0$}; ii) it contains no incoming waves from asymptotic infinity, neither physical nor coordinate waves; and iii)
it satisfies the consistency condition that the cones of this effective causality always lie inside the fundamental Minkowski cones, i.e.\ $[g] \leq [\eta]$; we have discussed that Penrose's property does not pose a fundamental obstruction to such construction. Concentrations of matter thus deform the causality in their surroundings ($g$) and produce delays (never advances) with respect to the more fundamental causality $\eta$. This complies with the original special relativity idea that nothing can travel faster than the speed of light.

In standard situations this gravitational theory is phenomenologically indistinguishable from GR and can be thought of as just an alternative interpretation. This is true because for the geometric and the field-theoretical paradigms, the harmonic background just amounts to the selection of a particular gauge in which to perform the calculations. However, by granting these conditions a physical reality (not merely based on a gauge choice), we obtain several very interesting bonuses. The first one is a possible explanation of the observational fact that matter is universally attractive in all standard situations.
By inverting the logical arrow of the VBL reasoning, we have argued that this fundamental causality hypothesis automatically implies, for consistency, that matter fields in weak field situations must satisfy a background null energy condition, in other words: that in all standard situations, matter must be universally attractive.

Our proposal has other advantages with respect to the geometric and the field-theoretical paradigms. We highlight specifically three points: i) As already observed by 
Rosen~\cite{Rosen1940a}, introducing a background metric $\eta$ allows to construct a SET for the gravitational field, rather than just a pseudo-tensorial quantity. Moreover, having a unique prescription for the gravitational field $h_{\mu\nu}$, it does not suffer from the interpretation problem it has in the field-theoretical paradigm, where the gravitational SET is not gauge-invariant. 
ii) The total fixing of $h_{\mu\nu}$ amounts to an unambiguous identification of the effective metric and the background geometry. Then, a curve defined over $g$ has a unique correspondence with respect to $\eta$. Thus, the geodesic trajectory, for example, of a planet around the Sun has a perfectly defined meaning for an observer who only perceives the background metric. 
iii) The bi-metrical theory is less isolated in theory space corresponding, for example, to the low-mass limit of contemporary massive-gravity theories. Beyond these concrete points, in
more general terms we expect this paradigm might simplify many conceptual issues that appear within the standard paradigms.

In any case, the most important asset of the harmonic background paradigm lies in the possible extensions from GR that it can accommodate, from Rosen-like modifications~\cite{Rosen1973} to contemporary massive gravity theories~\cite{deRham2014,Hinterbichler2012}, to name just the most obvious. In this context, the present authors proposed several years ago~\cite{Barceloetal2011} that a possible way of solving gravitational collapse singularities is by considering that gravity ``switches off'' when matter reaches Planckian densities (see also ~\cite{Barceloetal2015,Barceloetal2016}). This idea was originally motivated by condensed-matter models of gravity, in which hierarchies between effective and more fundamental levels of description emerge naturally.\footnote{It might be interesting to note that a bi-metric effective gravity with a flat background very similar to Logunov's RTG is spontaneously realized in $^3$He-A~\cite{volovik2003universe}, see the discussion in~\cite{jannes2012cosmological}. Other scenarios of emergent gravity are also possible, which do not necessarily depend on a flat background, see e.g.~\cite{klinkhamer2019tetrads,volovik2023acoustic}.
} We then argued that, if ``real'' gravity shows a similar emergent hierarchy, there might exist some energy scale at which the effective metric disappears as an order parameter of the system, leaving only the background causality to take control of the physics. From the point of view of such emergent-gravity scenarios, let us also mention that in a generic Lorentz-invariant theory of a tensorial field $h_{\mu\nu}$, without imposing any gauge symmetries, one recovers precisely the De Donder or harmonic condition (\ref{Eq:DonderGauge}) as a natural restriction on the space of solutions~\cite{Barceloetal2021}. Thus, starting from a theory without gauge symmetries, one deduces a dynamical theory for gravity equivalent to GR when written in a particular physical gauge, precisely corresponding to the harmonic background argued for in this paper.  

There are many subtleties still to be analyzed within the harmonic background paradigm, mostly in its extension to the full non-linear regime. 
We hope this paper serves to clear up the VBL/Gao-Wald controversy and to stimulate further analyses and
alternative ways of thinking about gravity.

\begin{acknowledgements}
The authors would like to thank Gerardo García-Moreno for enlightening discussions on the subject, as well as Stefano Liberati and Grigory Volovik for useful comments on the manuscript. 
Financial support was provided by the Spanish Government through projects PID2020-118159GB-C43 and PID2020-118159GB-C44 (with FEDER contribution), and by the Junta de Andaluc\'{\i}a through project FQM219. 
C.B. acknowledges financial support from the grant CEX2021-001131-S funded by MCIN/AEI/ 10.13039/501100011033.
\end{acknowledgements}

\bibliography{bi-metricity}

\end{document}